\documentstyle[prl,array,aps,twocolumn,epsf]{revtex}
\begin{document}
\draft
\twocolumn[\hsize\textwidth\columnwidth\hsize\csname @twocolumnfalse\endcsname
\title{Theory of Ferromagnetism in Ca$_{1-x}$La$_{x}$B$_6$}
\author{Shuichi Murakami$^{1}$ \cite{author}, Ryuichi Shindou$^{1}$, 
Naoto Nagaosa$^{1,2}$, 
and Andrei~S.~Mishchenko$^{2,3}$}
\address{$^{1}$Department of Applied Physics, University of Tokyo,
Bunkyo-ku, Tokyo 113-8656, Japan\\
$^{2}$Correlated Electron Research Center, Tsukuba 305-0046, Japan\\
$^{3}$RRC `Kurchatov Institute',123182, Moscow, Russia}
\date{\today}
\maketitle
\begin{abstract}
Novel ferromagnetism in Ca$_{1-x}$La$_{x}$B$_6$ is studied in terms 
of the Ginzburg-Landau theory
for excitonic order parameters, taking into account symmetry of the
wavefunctions. 
We found that the minima of the free energy break both inversion and 
time-reversal symmetries, while the product of these two remains
preserved. This explains various novelties of the ferromagnetism and 
predicts a number of magnetic properties, including the 
magnetoelectric effect, which can be tested experimentally.
\end{abstract}
\vskip2pc]

\narrowtext
\renewcommand{\labelenumi}{(\roman{enumi})}
 
Novel ferromagnetism in Ca$_{1-x}$La$_{x}$B$_6$ \cite{young} has been the 
subject of extensive studies due to its high Curie temperature ($\sim$ 600K) 
in spite of a small moment (0.07$\mu_{\rm B}$/La) and lack of partially 
filled  $d$- or $f$- bands. However, its mechanism still remains
controversial.  Ceperly suggested that this is a 
first example of the ferromagnetic phase of a dilute electron gas 
\cite{ceperley}. 
An alternative explanation \cite{zra,bv} is based on the excitonic state 
\cite{hr} of the parent material CaB$_6$. 
The parent compound CaB$_6$ has a cubic structure, and some band structure
calculations \cite{hy,mcpm} predict a small overlap of 
the conduction and valence bands at the X-points. Matrix elements of the
dipole moments vanish between these states at the X-points;
the dielectric constant is not 
enhanced even when the band gap collapses. This makes the excitonic 
insulator a plausible candidate for the ground state of CaB$_6$ \cite{zra}. 
By La-doping the extra electrons are doped into 
this excitonic insulator, and it has been found in the mean-field 
approximation that these electrons are perfectly spin-polarized \cite{zra}. 
Nevertheless, there is so far no conclusive explanation for the magnitude of 
the moment, much smaller than the electronic moment doped by La. 
Furthermore, the correlation between the La-doping concentration and the 
ferromagnetic moments are questioned by recent experiments \cite{takagi}. 
Several experiments suggest that the ferromagnetism is not a bulk phenomenon, 
and occurs only in the thin film sample 
\cite{terashima}, or near the surface as evidenced by an electron spin 
resonance (ESR) experiment
\cite{kunii}.

In this letter we study in depth the symmetry properties of the excitonic
state in CaB$_6$ in terms of the Ginzburg-Landau (GL) theory, and 
propose a possible scenario for the novel ferromagnetism.
We classify possible states in terms of the magnetic point groups.
The idea is that CaB$_6$ is a triplet excitonic insulator \cite{zra}
with broken time-reversal ($R$)  and inversion ($I$) symmetries, 
while their product $RI$ 
is kept intact \cite{zra}. 
This means that CaB$_6$ is an antiferromagnet (AF),
and the ferromagnetism is induced by the magneto-electric (ME) 
effect, and the La-doping and/or the surface works mainly as a source of 
the electric field. 
We emphasize that the present theory of ferromagnetism 
is different from those in 
\cite{zra,bv}, though all these theories are based on exciton
condensation.

This scenario of exciton condensation is justified when the band
structure has a small overlap/gap at the X points.
It is, however, still in controversy whether it is the case. 
Contrary to the 
LDA calculation \cite{hy,mcpm}, 
more recent LDA$+$GW calculation \cite{tromp}
suggested a large gap of $0.8{\rm eV}$, too large compared with 
the energy scale for exciton binding energy $(\sim 600{\rm K})$.
An angle-resolved photoemission spectrum (ARPES) \cite{ARPES} also shows
a large gap of order 1eV, and is in good accordance with the GW result.
Nevertheless, there are still enough evidences for believing a small gap
or a small overlap. First, another recent GW calculation \cite{kino}
shows a small overlap at the X points. Second, de Haas-van Alphen 
measurement \cite{hall} finds Fermi surfaces of both electrons and 
holes.
Third, the ARPES experiment is surface-sensitive, and one
can argue that the ARPES result \cite{ARPES} of the large gap might 
be attributed to surface effect.  
Fourth, measurements of X-ray scattering and  Raman scattering 
in CaB$_6$ \cite{udagawa} 
shows an anomaly at $600{\rm K}$, below which
the compound is determined to be tetragonal. 
This indicates that the material has some ordering and symmetry-breaking
even in the parent compound. 
This ordering is antiferromagnetic, as indicated in 
$\mu$SR measurements \cite{ohishi}.
Therefore, a large gap is unlikely, which prohibits any instability and 
sensitivity to the small amount of doping.
Thus, although there is still no concensus on a size of the gap, the
excitonic
insulator still remains one of the most promising candidates for the novel
ferromagnetism.

Let ${\bf Q}_{x},\ {\bf Q}_{y},\ {\bf Q}_{z}$
denote the three X points $(\pi,0,0),(0,\pi,0),(0,0,\pi)$
in the cubic Brillouin zone. 
When excitons are formed, excitonic order parameters 
$\langle b_{{\bf k}\alpha}^{\dagger}a_{{\bf k'}\beta}\rangle$
will have nonzero values, where $b_{{\bf k}\sigma}$ and 
$a_{{\bf k}\sigma}$ are annihilation operators of electrons with spin 
$\sigma$ at the conduction and the valence bands respectively \cite{hr}.
Because the excitonic instability occurs only in the vicinity of the 
three X points, we make the following assumptions;
\begin{enumerate}
 \item The order parameters are ${\bf k}$-independent near the X points.
 \item The order parameters connecting different X points are neglected
 \cite{bv}.
\end{enumerate}
As a result we keep only the order parameters 
$\eta({\bf Q}_{i})_{\alpha\beta}=
\langle b_{{\bf Q}_{i}\alpha}^{\dagger}a_{{\bf Q}_{i}\beta}\rangle$.
The cubic (${\rm O}_{\rm h}$)
symmetry of CaB$_6$ restricts a form of the GL free energy
$\Phi$, as is similar to the GL theory for 
unconventional superconductors \cite{vg,su}.

At the X points, the ${\bf k}$-group has a tetragonal (${\rm D}_{\rm 4h}$)
symmetry, and the conduction and the valence band states
belong to ${\rm X}_{3}$ and ${\rm X}^{\prime}_{3}$ representations, 
respectively \cite{hy,mcpm}. Without the spin-orbit coupling, 
the order parameter $\eta({\bf Q}_{i})_{\alpha\beta}$ belongs to 
${\rm X}_{3} \times {\rm X}_{3}^{\prime}={\rm X}^{\prime}_{1}$.
From now on we shall take the spin-orbit coupling into account. 
Then, the point-group 
transformation is accompanied by the spin rotation, and the representation 
of $\eta$ will be altered. 

Let us consider the triplet channel for the excitons, because the 
exchange interaction usually favors the triplet
 compared with the singlet \cite{zra}.
Then the spin-1 representation of the rotation group 
will be multiplied, and we get 
${\rm X}_{4}^{\prime}+{\rm X}_{5}^{\prime}$ in the ${\rm D}_{{\rm 4h}}$
group. Explicitly, for the triplet excitons at the ${\bf Q}_{z}$, 
the $S_{z}=0$ component $\eta_{\downarrow\downarrow}-\eta_{\uparrow\uparrow}$
obeys the ${\rm X}_{4}^{\prime}$ representation, while the $S_{z}=\pm 1$
components 
$\eta_{\uparrow\downarrow},\ \eta_{\downarrow\uparrow}$ 
obey the 
${\rm X}_{5}^{\prime}$.
Here the spin-quantization axis is taken to be $+z$-direction.
Since we are taking into account 
the spin-orbit coupling, 
these up- and down-spins
should be interpreted as pseudospins \cite{su}.

When we consider the three X points, the cubic symmetry is restored
since the order parameters have a wavevector ${\bf q}={\bf 0}$.
The order parameters have nine components in total, which
can be classified into irreducible representations of 
${\rm O}_{\rm h}$. They split 
into three 3-dimensional
representations: ${\rm \Gamma}_{\rm 15}+{\rm \Gamma}_{\rm 15}+
{\rm \Gamma}_{\rm 25}$.
Let us call basis functions as $\eta_{i}({\rm \Gamma}_{15},1)$,
$\eta_{i}({\rm \Gamma}_{15},2)$ and $\eta_{j}({\rm \Gamma}_{25})$,
where $i=x,y,z$ and $j=x(y^2 -z^2),y(z^2-x^2),z(x^2-y^2)$.
These basis functions transform like a vector $(x,y,z)$ for
$\Gamma_{15}$ and like $(x(y^2 -z^2),y(z^2-x^2),z(x^2-y^2))$ for 
$\Gamma_{25}$.
By judicious choice of phase for basis functions, they can be 
made to transform under the time-reversal $R$ as complex conjugation,
and they are odd under the inversion $I$; i.e.
$ I \eta = - \eta $, and $ R \eta = \eta^* $.
The basis functions are given as
\begin{eqnarray}
&& \eta_{z}({\rm \Gamma}_{15},1)=
\frac{{\rm i}}{\sqrt{2}}
(\eta_{\downarrow\downarrow}({\bf Q}_{z})-
\eta_{\uparrow\uparrow}({\bf Q}_{z})), \\
&& \eta_{z}({\rm \Gamma}_{15},2)=
-\frac{1}{2}\{
{\rm i}(\eta_{\uparrow\downarrow}({\bf Q}_{y})+
\eta_{\downarrow\uparrow}({\bf Q}_{y}))\nonumber \\
&&\makebox[3mm]{}
+(\eta_{\uparrow\downarrow}({\bf Q}_{x})-
\eta_{\downarrow\uparrow}({\bf Q}_{x}))
\}
, \\
&& \eta_{z(x^2 -y^2)}({\rm \Gamma}_{25})=
\frac{1}{2}\{
{\rm i}(\eta_{\uparrow\downarrow}({\bf Q}_{y})+
\eta_{\downarrow\uparrow}({\bf Q}_{y}))\nonumber \\
&&\makebox[3mm]{}
-(\eta_{\uparrow\downarrow}({\bf Q}_{x})-
\eta_{\downarrow\uparrow}({\bf Q}_{x}))
\}.
\end{eqnarray} 
Here, the spin-quantization axis in 
$\eta_{\alpha\beta}({\bf Q}_{i})$ is taken as $+i$-axis ($i=x,y,z$). 
Other components are obtained by cyclic permutation of $x,y,z$.

Let us now write down the GL free energy in terms of these order
parameters. The GL free energy $\Phi$ should be invariant under the 
elements of ${\rm O}_{\rm h}$ and under the time-reversal $R$.
We make two remarks helpful in writing down $\Phi$.
First, only even-order terms in {\boldmath$\eta$} are allowed
by the inversion symmetry. Second, owing to the time-reversal symmetry, 
the order of ${\rm Im}\mbox{\boldmath$\eta$}$ in each term should be even.
Thus, $\Phi$ is given up to quadratic order as
\begin{eqnarray}
 &&\Phi^{(2)}=A_{1}\sum_{i}({\rm Re}\ \eta_{i}(\Gamma_{15},1))^{2}
+A_{2}\sum_{i}({\rm Re}\ \eta_{i}(\Gamma_{15},2))^{2}
\nonumber \\
&&\makebox[.5mm]{}
+A_{3}\sum_{i}
{\rm Re}\ \eta_{i}(\Gamma_{15},1){\rm Re}\ \eta_{i}(\Gamma_{15},2)
+A_{4}\sum_{j}({\rm Re}\ \eta_{j}(\Gamma_{25}))^{2}\nonumber \\
&&\makebox[.5mm]{}
+B_{1}\sum_{i}({\rm Im}\ \eta_{i}(\Gamma_{15},1))^{2}
+B_{2}\sum_{i}({\rm Im}\ \eta_{i}(\Gamma_{15},2))^{2}
 \nonumber \\
&&\makebox[.5mm]{}+
B_{3}\sum_{i}
{\rm Im}\ \eta_{i}(\Gamma_{15},1){\rm Im}\ \eta_{i}(\Gamma_{15},2)
+B_{4}\sum_{j}({\rm Im}\ \eta_{j}(\Gamma_{25}))^{2} \label{phi2}
\end{eqnarray}
Therefore, as the temperature is lowered, one of the following states 
will be realized;
\renewcommand{\labelenumi}{(\Roman{enumi})}
\begin{description}
 \item[(A)] ${\rm Re}\ \mbox{\boldmath$\eta$}(\Gamma_{15},1) = c 
        {\rm Re}\ \mbox{\boldmath$\eta$}(\Gamma_{15},2)$ $\ne {\bf 0}$,
 \item[(B)] 
${\rm Re}\ \mbox{\boldmath$\eta$}(\Gamma_{25})\neq 0$, 
 \item[(C)] ${\rm Im}\ \mbox{\boldmath$\eta$}(\Gamma_{15},1) = c
        {\rm Im}\ \mbox{\boldmath$\eta$}(\Gamma_{15},2)$ $\ne {\bf 0}$,
 \item[(D)] ${\rm Im}\ \mbox{\boldmath$\eta$}(\Gamma_{25})\neq {\bf 0}$, 
\end{description}
where $c$ is a nonzero constant.
Therefore, a condensation of excitons in $\Gamma_{15}$ and in
$\Gamma_{25}$ do not occur simultaneously. 
In each case among (A)-(D), 
all directions of the vector {\boldmath$\eta$} are degenerate, 
and this degeneracy will be lifted in the quartic order, 
as we shall see later. All these states are accompanied
by a lattice distortion. 
This distortion, however, 
is expected be rather small because it couples to the order 
parameter in the quadratic order, not linear order.
Indeed the tetragonal distortion detected by X-ray scattering
is as small as $0.03\%$ \cite{udagawa}.

The states (A),(B) preserve time-reversal 
symmetry and thus nonmagnetic. From symmetry consideration,
this implies that neither the ME nor the 
piezomagnetic (PM) effect will be observed 
\cite{birss}.
Roughly speaking these states are far from showing ferromagnetism.

To study which state is realized,
we should know the coefficients $A_{i}$, $B_{i}$ in eq.(\ref{phi2}), 
which are related to the  
following matrix elements for the exchange
interaction by 
the Hartree-Fock approximation \cite{hr}.
\begin{equation}
\int{\rm d}{\bf x}\int{\rm d}{\bf x'}
\phi_{{\bf Q}_{i}}^{c}({\bf x})
\phi_{{\bf Q}_{j}}^{c'}({\bf x})\frac{e^2}{\varepsilon|{\bf x}-{\bf x'}|}
\phi_{{\bf Q}_{j}}^{c''}({\bf x'})\phi_{{\bf Q}_{i}}^{c'''}({\bf x'})
,
\end{equation}
where
$\phi_{{\bf Q}_{i}}^{c}({\bf x})$ is Bloch wave function of the $c$-band
($c$ = a,b) at ${\bf k}={\bf Q}_{i}$, $\varepsilon$ is a
dielectric constant.
In the 
calculation of these matrix elements, we neglect the spin-orbit coupling.
To calculate $\phi_{{\bf Q}_{i}}^{c}({\bf x})$ we proceed as follows.
We define
$\phi_{{\bf Q}_{i},{\rm X}_{3}}^{\rm{B:p}}$ and
$\phi_{{\bf Q}_{i},{\rm X}_{3}'}^{\rm{B:p}}$  as
linear combinations of $p$-orbitals of B with 
${\rm X}_{3}$ and ${\rm X}_{3}'$, respectively, and 
$\phi_{{\bf Q}_{i},{\rm X}_{3}'}^{\rm{Ca:d}}$ as a
linear combination of $d$-orbitals of Ca with 
${\rm X}_{3}^{\prime}$. 
Then, $\phi_{{\bf Q}_{i}}^{a}({\bf x})$ is defined as
a bonding orbital of 
$\phi_{{\bf Q}_{i},{\rm X}_{3}'}^{\rm{B:p}}$ and 
$\phi_{{\bf Q}_{i},{\rm X}_{3}'}^{\rm{Ca:d}}$,
and  $\phi_{{\bf Q}_{i}}^{b}({\bf x})$ is equal to 
$\phi_{{\bf Q}_{i},X_{3}}^{\rm{B:p}}$, 
with proper normalization \cite{hy}. 
We evaluated the coefficients in (\ref{phi2}), and found
that even if coupling terms between the three X-points
are included, the coefficients for 
the imaginary parts of the order parameters are smaller than
those for the real parts .
Therefore, the states (C) and (D) would be favorable than 
(A) or (B). Thus, from now on we shall concentrate on (C) and (D)
\cite{hr-comment}.

The states (C) and (D) break both the inversion and 
the time-reversal symmetries; they are magnetic states.
A crucial observation is that they preserve the $RI$ symmetry \cite{zra},
leading to many interesting consequences listed below.
(I) The $RI$ symmetry prohibits a uniform magnetic moment.
Dzyaloshinskii used this symmetry to explain
why  weak ferromagnetism is present in $\alpha$-Fe$_2$O$_3$ while not in
Cr$_2$O$_3$ \cite{dzy1}.
Thus the states (C) and (D) are antiferromagnetic. This agrees with 
the result of the $\mu$SR measurement \cite{ohishi} 
with a moment of $0.0039\mu_{{\rm B}}/{\rm mol}$. 
Note that the magnetic unit cell is identical
with the original unit cell. Thus,
no extra Bragg spots appear below the AF transition.
(II) The $RI$ symmetry also prohibits 
the PM effect. 
This follows because a stress cannot break the $RI$ 
symmetry which impedes ferromagnetism.
Nevertheless, a {\it gradient} of stress can break this 
symmetry and will induce ferromagnetism.
We note in passing that if intervalley excitons 
condense, i.e. the assumption (ii) is violated, the PM effect can occur.
(III) The $RI$-invariance results in the linear ME effect, as 
observed in Cr$_2$O$_3$ \cite{dzy2}. 
This occurs because an external electric field $\bf E$
breaks this $RI$ symmetry and enables ferromagnetism \cite{dzy2}.
An electric field $\bf E$ belongs to the ${\rm \Gamma}_{15}$, 
and couples linearly with the order parameters in ${\rm \Gamma}_{15}$ as
\begin{equation}
 \delta\Phi =
-C_{1}{\bf E}\cdot{\rm Re}\ \mbox{\boldmath$\eta$}(\Gamma_{15},1)
-C_{2}{\bf E}\cdot {\rm Re}\ \mbox{\boldmath$\eta$}(\Gamma_{15},2).
\end{equation}
Here the imaginary parts of the order parameters are absent
due to invariance of $\delta\Phi$ under time-reversal.
Thus, in the presence of ${\bf E}$, both the real and imaginary 
parts of the order parameters acquire nonvanishing values, resulting 
in a uniform moment.
As for the $\Gamma_{25}$, similar effect can be found.
(IV) The optical non-reciprocal (NR) effect in reflection 
is predicted to occur, as is similar to Cr$_2$O$_3$
\cite{kppg}. 
This effect will occur only below the N{\'e}el temperature.
Generally, $RI$-invariant materials
with broken $R$- and $I$-symmetry will exhibit the NR effect \cite{kppg}.

Let us consider quartic order terms $\Phi^{(4)}$ in the GL free energy, 
which lifts the degeneracy in the direction of {\boldmath$\eta$}.
We do not write down its lengthy formula here \cite{msnm}. 
By minimizing $\Phi^{(2)}+\Phi^{(4)}$,
we find four possibilities;
\begin{description}
\item[(C-1)] ${\rm Im}\ \mbox{\boldmath$\eta$}(\Gamma_{15},1) = c_1 
        {\rm Im}\ \mbox{\boldmath$\eta$}(\Gamma_{15},2) = (0,0,c_2) $,
\item[(C-2)] ${\rm Im}\ \mbox{\boldmath$\eta$}(\Gamma_{15},1) = c_1
        {\rm Im}\ \mbox{\boldmath$\eta$}(\Gamma_{15},2)= (c_2,c_2,c_2) $,
\item[(D-1)] ${\rm Im}\ \mbox{\boldmath$\eta$}(\Gamma_{25}) = (0,0,c)$,
\item[(D-2)] ${\rm Im}\ \mbox{\boldmath$\eta$}(\Gamma_{25})=(c,c,c)$,
\end{description}
where $c$'s are constants.
The  direction of the lattice distortion is 
tetragonal in (C-1) and (D-1) and is trigonal 
in (C-2) and (D-2).
Magnetic point groups $G$ for these states are
(C-1) 4/${\rm m}'$mm, (C-2) $\bar{3}'{\rm m}$, (D-1) 
$4'/{\rm m}'{\rm m}'{\rm m}$, 
(D-2) $\bar{3}'{\rm m}'$.
In each case among (C-1)-(D-2), there are two types of degenerate AF
domains, when we fix the axis of tetragonal or trigonal distortion.
Since there are three and four choices of axes for tetragonal and trigonal
cases, respectively, total degeneracy is six in (C-1)(D-1) and eight in 
(C-2)(D-2), which is equal to an order of the quotient group 
$({\rm O}_{\rm h}\times\{ E,R \})/G$.
We can draw some analogies with anisotropic superconductivity (SC).
The order parameters {\boldmath$\eta$} of triplet excitons correspond 
to the d-vector in triplet SC. It is nevertheless misleading to 
look for SC counterparts of our phases (C-1)-(D-2), because our order 
parameters are confined in the neighborhood of the three X points.
They are triplet and even functions in ${\bf k}$, which never occurs 
in the SC.

Information for an ME effect can be obtained from ref.~\cite{birss}.
When we write a bilinear term of ${\bf H}$ and ${\bf E}$ in the free
energy as 
$\alpha_{ij}H_{i}E_{j}$, the property tensor $\alpha_{ij}$ is 
\begin{eqnarray*}
&&\mbox{(C-1)} 
\left(
\begin{array}{ccc}  0 & \alpha_{1} &0 \\
  -\alpha_{1}&0&0\\
  0&0&0       
\end{array}
\right), \ 
\mbox{(C-2)}
\left(
\begin{array}{ccc}
  0&\alpha_{1}&-\alpha_{1}\\
  -\alpha_{1}&0&\alpha_{1}\\
  \alpha_{1}&-\alpha_{1}&0       
\end{array}
\right), \\
&&\mbox{(D-1)}
\left(
\begin{array}{ccc}
  0 & \alpha_{1} &0\\
  \alpha_{1}&0& 0\\
  0&0&0       
\end{array}
\right), \ 
\mbox{(D-2)}
\left(
\begin{array}{ccc}
  \alpha_{1}&\alpha_{2}&\alpha_{2}\\
  \alpha_{2}&\alpha_{1}&\alpha_{2}\\
  \alpha_{2}&\alpha_{2}&\alpha_{1}       
\end{array}
\right).
\end{eqnarray*}
These explicit forms of $\mbox{\boldmath$\alpha$}$
contain information on 
a direction of magnetization ${\bf M}$ under an electric field 
${\bf E}$ and that of polarization ${\bf P}$
under a magnetic field ${\bf B}$.
In particular, in (C-1) and (C-2), ${\bf M}\bot{\bf E}$ and 
${\bf P}\bot{\bf B}$ always hold.
Thus, measurement of the ME effect with a single crystal of 
CaB$_6$ will manifest which of these four is realized.
Experimentally, domain structure of the AF 
develops, and the sign of 
$\mbox{\boldmath$\alpha$}$ is reversed when the staggered magnetization is 
reversed. In measurement of the ME effect, 
the domain structure can be aligned by  
a magnetoelectric annealing, in which the sample is 
cooled under both electric and  magnetic fields.
Domain boundaries between the two AF domains can exhibit interesting 
properties. As in boundaries between two SC domains with 
broken time-reversal symmetry \cite{su}, localized current and 
magnetic moment are induced near the boundary. In the present case 
it is interpreted as the ME effect.
%

We remark on the recent results of 
X-ray scattering and Raman scattering \cite{udagawa},
which strongly support our theory. They show 
a tetragonal distortion below 600K, which indicates that (C-1) or (D-1)
is the case. Furthermore, 
inversion-symmetry breaking does not appear in the Raman 
spectrum \cite{udagawa2}, which is consistent with (C) or (D), but not
with (A) or (B). Thus, (C-1) and (D-1) are the only possibilities 
totally consistent with \cite{udagawa}.

The ferromagnetism in the thin-film CaB$_6$ is interpreted as caused by an 
electric field between vacuum and the substrate. 
For the powder experiment and the La-doping experiment, 
the explanation is more delicate. We believe that the carriers by La-doping 
are trapped by impurities/defects and create local electric fields.  
Therefore, an internal electric field 
and/or a gradient of a strain has a random direction, 
and hence the magnetic moment is induced locally due to this mechanism. 
At first sight they appear to cancel with each other,
giving zero or quite small uniform magnetization, which 
contradicts with the experiments.
However the random direction of the frozen electric field and 
the external magnetic field determine the domain structure of the 
AF; the free energy to be minimized is
composed of (i) the 
energy gain due to the magnetization in the external magnetic field 
and (ii) the elastic energy loss of the spatial change of the order parameter. 
Therefore, the pinning of the domain-wall motion leads to the 
hysteresis behavior, which is regarded as the experimental signature of the 
``ferromagnetism". The detailed study of the magnetization curve
including its dependence on the sweeping time,
however, can distinguish it from the real ferromagnetism. 

Other peculiarities of Ca$_{1-x}$La$_x$B$_6$ can also be explained as well.
The high Curie temperature $(\sim 600{\rm K})$ is nothing but 
a N{\'e}el temperature of the parent compound CaB$_6$, and is not
contradictory 
with a tiny magnetic moment. A rather narrow range ($x\lesssim 0.01$) 
of La-doping 
allowing ferromagnetism is attributed to fragility of excitonic order 
by a small amount of impurities \cite{sk,z}.
Moreover, our scenario is also consistent with the experimental results
that deficiency in Ca sites \cite{morikawa} or
doping of divalent elements like Ba \cite{young} or Sr \cite{ott} 
induces ferromagnetism.
It is also confirmed numerically by a supercell approach that
imperfections and surfaces can induce a local moment \cite{md}.
It is hard to explain them within the spin-doping scenario \cite{zra}.
Furthermore, strangely enough,
it is 
hard to find a correlation between magnetism and electrical 
resistivity, as seen in magnetizaion \cite{morikawa} and in 
nuclear magnetic resonance \cite{gavilano}. This novelty is
a natural consequence of our scenario; electrical 
resistivity is mainly due to doped carriers, while the 
magnetization is due to local lattice distortion and/or electric field.

The ESR experiments by Kunii \cite{kunii}
also support the above scenario. The ESR data show that in 
a disk-shaped Ca$_{0.995}$La$_{0.005}$B$_6$,
the magnetic moment only exists within the surface layer of
$\sim 1.5\mu {\rm m}$ thick.
Furthermore, the moment ${\bf M}$ does not orient in the direction
of ${\bf H}$, i.e. it feels strong magnetic anisotropy to keep the 
moment within the disk plane. This might be due to the long-range 
dipolar energy, and not due to the above scenario.
Nevertheless, it is unlikely that the long-range dipolar energy 
causes such a strong anisotropy.
This point requires further investigation.

Let us, for the moment, assume that this strong anisotropy 
is mainly caused by the exciton condensation and the ME mechanism.
Since this electric field should be
perpendicular to the plane, the strong easy-plane anisotropy 
parallel to the surface implies that ${\bf M}\bot {\bf E}$.
Therefore, among the four cases, (C-1) or (C-2) are compatible.
Together with the result of Raman scattering, (C-1) is the most promising
candidate for CaB$_6$.

We mention here a role of the spin-orbit coupling. In the limit of  
zero spin-orbit coupling, the states (C) and (D) become degenerate, 
and the quartic-order terms in the GL free energy do not lift this
degeneracy. 
Sixth-order terms will lift it, and a resulting state
will belong to either 4/${\rm m}'$mm or  
$4'/{\rm m}'{\rm m}'{\rm m}$. Both of them still lead to the ME 
effect in the absence of the 
spin-orbit coupling;
this ME effect must be generated from an orbital motion.
Thus, the AF state in CaB$_6$ has an orbital nature as 
well as a spin nature. With spin-orbit interaction, these two
are inseparably mixed together.

In conclusion, we have studied the symmetry properties of the 
excitonic state in CaB$_6$, and found that the 
triplet excitonic state with broken time-reversal and inversion symmetries 
offers a natural explanation, in terms of the ME effect, 
for the novel ferromagnetism
emerging in La-doping or thin-film fabrication. 
This scenario can be tested experimentally by
measurements of the ME effect and
the optical non-reciprocal effect in single crystal of the parent compound
CaB$_6$.

The authors thank helpful discussion with Y.~Tokura, H. Takagi, Y. Tanabe
and K.~Ohgushi. We acknowledge support by 
Grant-in-Aids
from the Ministry of Education, Culture, Sports, Science and Technology.


\begin{thebibliography}{99}
\bibitem{hr-comment}
The pure imaginary order parameters in this letter correspond to
Class II and III in \cite{hr}, which break time-reversal symmetry. 
Likewise the real parts correspond to 
Class I and IV in \cite{hr}, with time-reversal symmetry being preserved.
Thus our observation that the order parameters become pure imaginary 
coincides with the result in \cite{hr} that the mixture of Classes II 
and III is the most 
favorable state in the presence of spin-orbit coupling, provided the 
electron-phonon coupling is not so strong.
\end{thebibliography}
\end{document}